\documentclass[final,5p,times,twocolumn]{elsarticle}
\usepackage{lineno,hyperref}
\modulolinenumbers[5]

\usepackage{amsmath,amsfonts}
\usepackage{algorithmic}
\usepackage{algorithm}
\usepackage{array}
\usepackage{subfig}
\usepackage{textcomp}
\usepackage{stfloats}
\usepackage{url}
\usepackage{verbatim}
\usepackage{graphicx}
\usepackage{color}
\usepackage{array}
\usepackage{graphicx}
\usepackage{graphicx}
\usepackage{siunitx}
\usepackage{subcaption}

\journal{Nano Communication Networks} % Replace with the actual Elsevier journal name

\begin{document}

\begin{frontmatter}

\title{Electrophoretic  Beamforming in Molecular Communication: Toward Targeted Extracellular Vesicle Delivery}
\author[ous]{Mohammad Zoofaghari \corref{cor1}}
\ead{mohzoo@ous-hf.no}

\author[ntnu]{Liv Cornelia Middelthon}
\ead{livcorneliam@gmail.com}

\author[ous]{Mladen Veletić }
\ead{mladen.veletic@ntnu.no }
\author[ntnu,ous]{Ilangko Balasingham}
\ead{ilangko.balasingham@ntnu.no}

\cortext[cor1]{Corresponding author}
\address[ous]{The Intervention Centre, Oslo University Hospital, 0372 Oslo, Norway}

\address[ntnu]{Department of Electronic Systems, Norwegian University of Science and Technology, 7491 Trondheim, Norway}

\begin{abstract}
Directing extracellular vesicles (EVs), such as exosomes and microvesicles, toward specific cells is an emerging focus in nanomedicine, owing to their natural role as carriers of proteins, RNAs, and drugs. EVs can be manipulated by external electric fields due to their intrinsic surface charge and biophysical properties.
This study investigates the feasibility of using extremely low-frequency electromagnetic fields to guide EV transport. A theoretical framework based on the Fokker–Planck equation was developed and numerically solved to model vesicle trajectories under time-harmonic drift. Computational simulations were conducted to systematically assess the influence of key electric field parameters—including phase, frequency, and intensity—on vesicle displacement and trajectory.
The findings demonstrate that frequencies below $5Hz$ combined with field strengths of $200-2000V/m$ can induce substantial directional control of EV motion. Moreover, enhanced directivity was achieved through the application of multi-component electric fields.
Overall, this work establishes a theoretical foundation for the external-field-based beamforming of nanoparticles within the framework of molecular communication.
 
\end{abstract}


\end{frontmatter}

%\linenumbers

% ==================
% Start of main text
% ==================

\section{Introduction}
\begin{figure}[!t]
\centering
\includegraphics[width=3.2in]{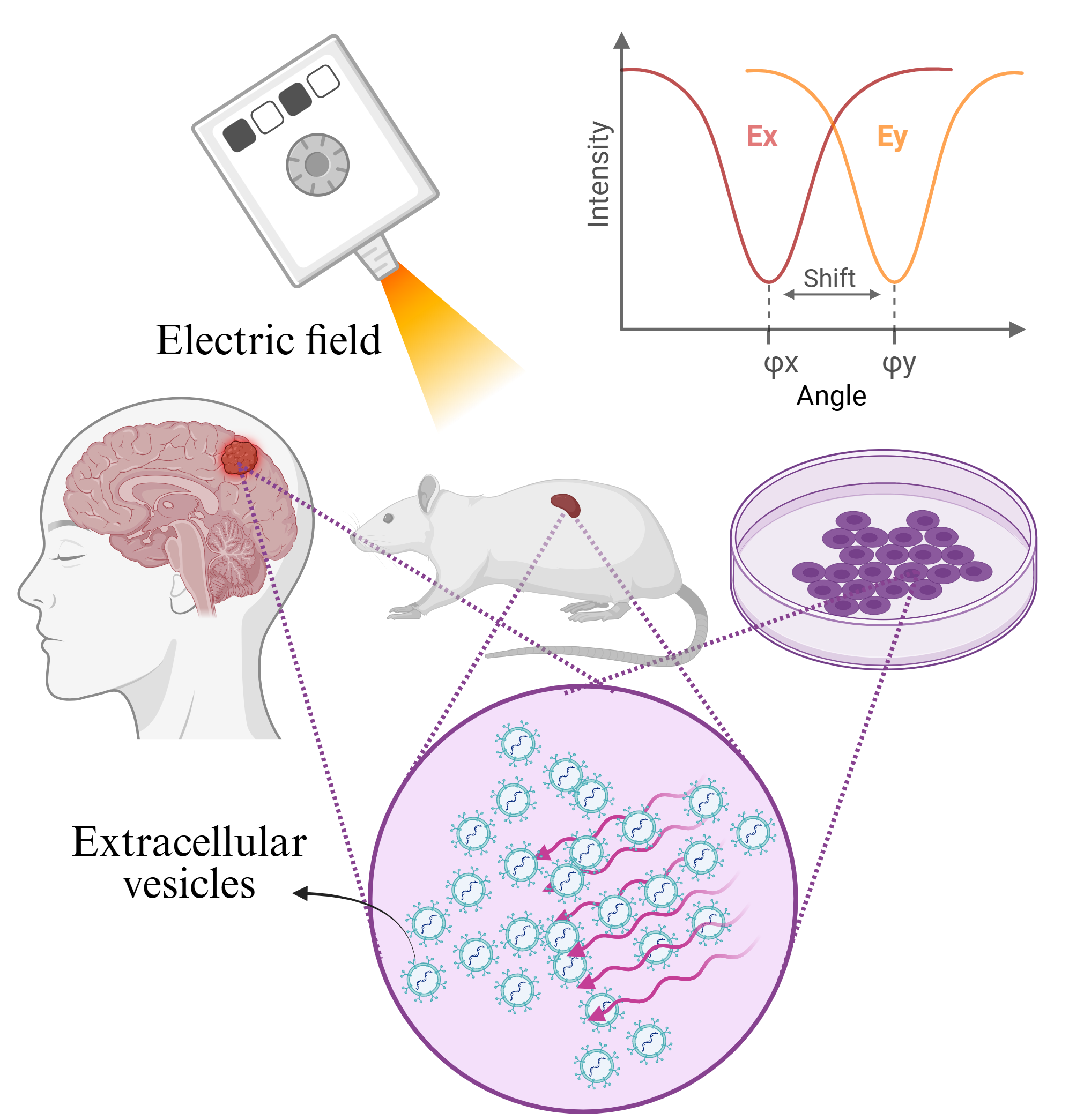}
\caption{Capability of electrophoretic beamforming in in vitro, in vivo, and clinical studies.}
\label{BF}
\end{figure}
Intrabody nanonetworks, composed of synthetic nanorobots designed for sensing and actuation, as well as synthetic or native cells such as neurons or immune cells that mediate intercellular communication, have been widely investigated in the literature in recent years \cite{gomez2025communicating}. Molecular communication has been introduced as a biocompatable and low-energy paradigm for nanonetworks; nevertheless, it continues to face considerable challenges relative to conventional RF-based systems. One of the limitations of MC systems is the lack of feasible mechanisms for molecular beamforming, which could enable high-precision drug delivery, reduction of off-target interference, improved resource efficiency, spatial multiplexing, dynamic adaptability, and enhanced network scalability. A molecular beamformer would allow steering of a directional molecular signal toward a specific population of nanomachines or cells.

Molecular beamforming enables switchable, spatially-targeted cell programming, including the on-demand release of siRNA/ASOs, CRISPR RNPs, PROTACs, cytokines, and other therapeutic agents; the modulation of signaling pathways; and the execution of logic gating at the cellular level.
Molecular beamforming also facilitates the scheduling of nanomachines for efficient information exchange. Real-time beam steering relies on external control signals—such as ultrasound, optical, magnetic, or electric fields—depending on the properties of the nanoparticles interacting with the applied field. Electrophoretic beamforming requires electric fields in conjunction with charged nanoparticles to enable drift, localization, and modulation of nanoparticle distributions under controlled field parameters, applicable to in vitro, in vivo, and human studies (Fig. \ref{BF}).
%Nanoparticles can serve as the mediators between the transmitter and the target cells by carrying nucleic acids, proteins, antibodies, or aptamers to enable precise cell-batch targeting. By redistributing the interaction of these nanoparticles—through external fields with tunable parameters—it becomes possible to switch selected subgroups of cells into new functional states.

\color{black}
%In recent years, targeted drug delivery (TDD) has advanced significantly, offering new strategies for improving the precision and efficiency of therapeutic interventions see Fig. \ref{fig:TDD}. 
%General approaches within TDD include using carriers that respond to specific biological markers, stimuli-responsive delivery systems, and the application of engineered particles or molecules capable of navigating biological barriers. %These methods are being explored across various medical contexts, including oncology, neurology, and cardiovascular medicine.

As a specific application, molecular beamforming can be utilized to steer therapeutic, stimuli-responsive extracellular vesicles (EVs).
%by synthetic implant cells toward glioblastoma cancer cells . Glioblastoma is an aggressive type of brain cancer that arises from mutations in glial cells [2]. Normally, glial cells support neurons and maintain the surrounding environment, but when mutated, they proliferate rapidly and form tumors. These tumors often infiltrate healthy brain tissue, making them particularly difficult to treat. Furthermore, the extracellular space within glioblastoma is both dense and structurally complex, which poses significant barriers to the delivery of therapeutic agents and reduces the effectiveness of conventional treatments.
Cells naturally release EVs, which are nanoscale, membrane-bound particles that mediate communication between cells by transporting molecular cargo such as mRNA, proteins, and lipids through the extracellular space (ECS) \cite{ref31}. Due to their biocompatibility and ability to carry therapeutic agents, EVs have emerged as promising candidates for targeted drug delivery. A central idea in this context is to exploit EVs not only as passive carriers but as actively steerable agents whose trajectories can be influenced after their release. Directing the motion of EVs through the ECS may enable more precise targeting of pathological cells, such as  glial cells, in in vitro and in vivo experiments (Fig. \ref{fig:TDD}). 
\subsection{Biological Background}
The human body relies on a highly complex and coordinated communication system to regulate its functions and maintain homeostasis. This system, often called biocommunication, involves the continuous exchange of information between cells, tissues, and organs. Through this exchange, the body can respond to changes in its internal and external environments, initiate repair processes, regulate growth, and carry out various physiological tasks. 
Biocommunication occurs across multiple levels of biological organization, from molecular interactions within cells \cite{10149035} to long-range signaling between distant tissues through the circulation system \cite{10922788}. Signals can take many forms, including molecular messengers, electrical impulses, and physical cues. Together, these mechanisms form an integrated network that enables precise and dynamic regulation of biological processes. This study concerns a specific form of molecular biocommunication mediated by EVs \cite{9446681}, which play a key role in intercellular signaling by transporting functional cargo between cells \cite{10032172}.

\begin{figure}[!t]
\centering
\includegraphics[width=3.5in]{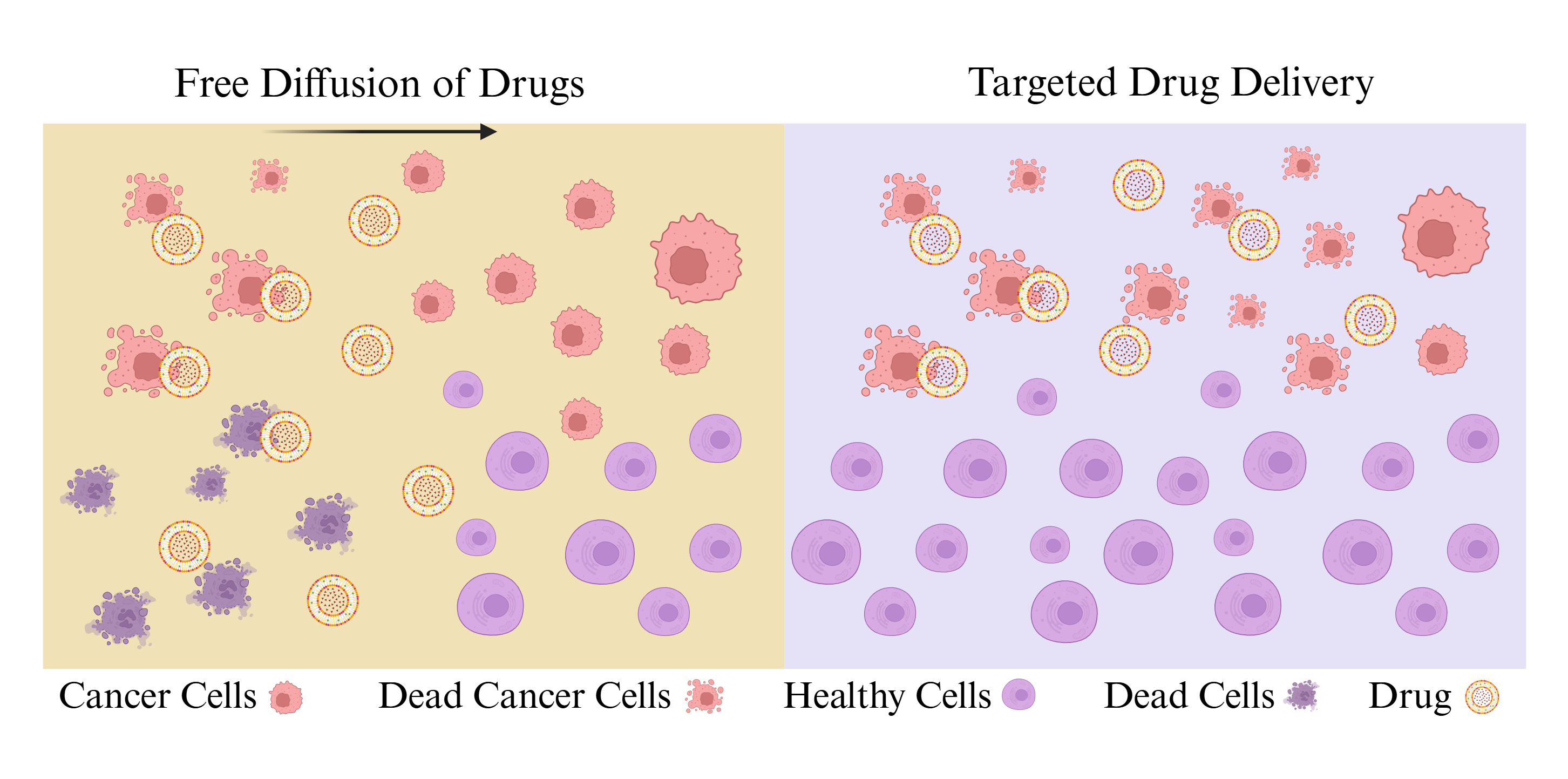}
\caption{Illustration of chemotherapy delivery to the cancer cells, comparing unassisted diffusion with controlled, targeted drift\cite{ref45}.}
\label{fig:TDD}
\end{figure}

Electromagnetic fields (EMFs) can interact with the physical properties of EVs, such as their surface charge, thereby influencing their drift behaviour. A significant consideration is to avoid methods that generate heat, radiation, or other harmful side effects. Extremely low frequency (ELF) EMFs, characterized by low energy and long wavelengths, are well-suited for biological applications requiring non-invasive modulation. Prior studies have shown that ELF EMFs can regulate both the release rate and the molecular composition of EVs in a frequency-dependent manner \cite{ref11}, highlighting their potential role in modulating EV-mediated communication. However, therapeutic application of EMFs in sensitive environments such as the brain presents several challenges, including ensuring adequate field penetration, maintaining directional precision, and minimizing off-target stimulation within the ECS. This study explores the feasibility of EVs beamforming to improve the precision and safety of targeted chemotherapy (see Fig. \ref{fig:TDD}), tissue regeneration or immunomodulation. To address these concerns, it is critical to design and rigorously evaluate field geometries that allow precise, safe, and effective control over EV transport.

A study by Wang et al. \cite{wang2023frequency} indicates that EMFs can influence both the release and composition of EVs, with important implications for fundamental biology and medical applications. Similarly, Wong et al. \cite{wong2022brief} demonstrate that directionally specific pulsed EMFs can stimulate EV release, highlighting potential scientific, medical, and commercial applications.
Previous studies on the behavior of EVs after their release into the ECS  have largely focused on diffusion, typically modeled using three-dimensional advection-diffusion equations \cite{rudsari-2022}. These models often account for key factors such as tortuosity, volume fraction, and fluid dynamics.
Moreover, EVs can be engineered to carry magnetic nanoparticles (e.g., iron oxide) or ions such as calcium, enabling their manipulation by external EMFs for targetted drug delivery applications \cite{wang2024engineering}.
This study examines the influence of ELF EMF on the likelihood of EVs reaching the receiver cell. This is achieved by evaluating the probability distribution of EV positions following their release from the transmitter. To model this behavior, the Fokker-Planck equation  is solved, incorporating a sinusoidal drift term as a simplified representation of the underlying transport mechanisms.
\section{System model}
This section presents the theoretical framework underlying the transport of EVs under the influence of ELF EMFs. It provides the essential physical and mathematical concepts needed to understand and model the behaviour of EVs in such fields. The proposed approach can be readily generalized to the beamforming of various types of nanoparticles in molecular communication, enabling efficient receiver scheduling.

\subsection{Transport Mechanisms}

\begin{figure}[!t]
\centering
\includegraphics[width=3.5in]{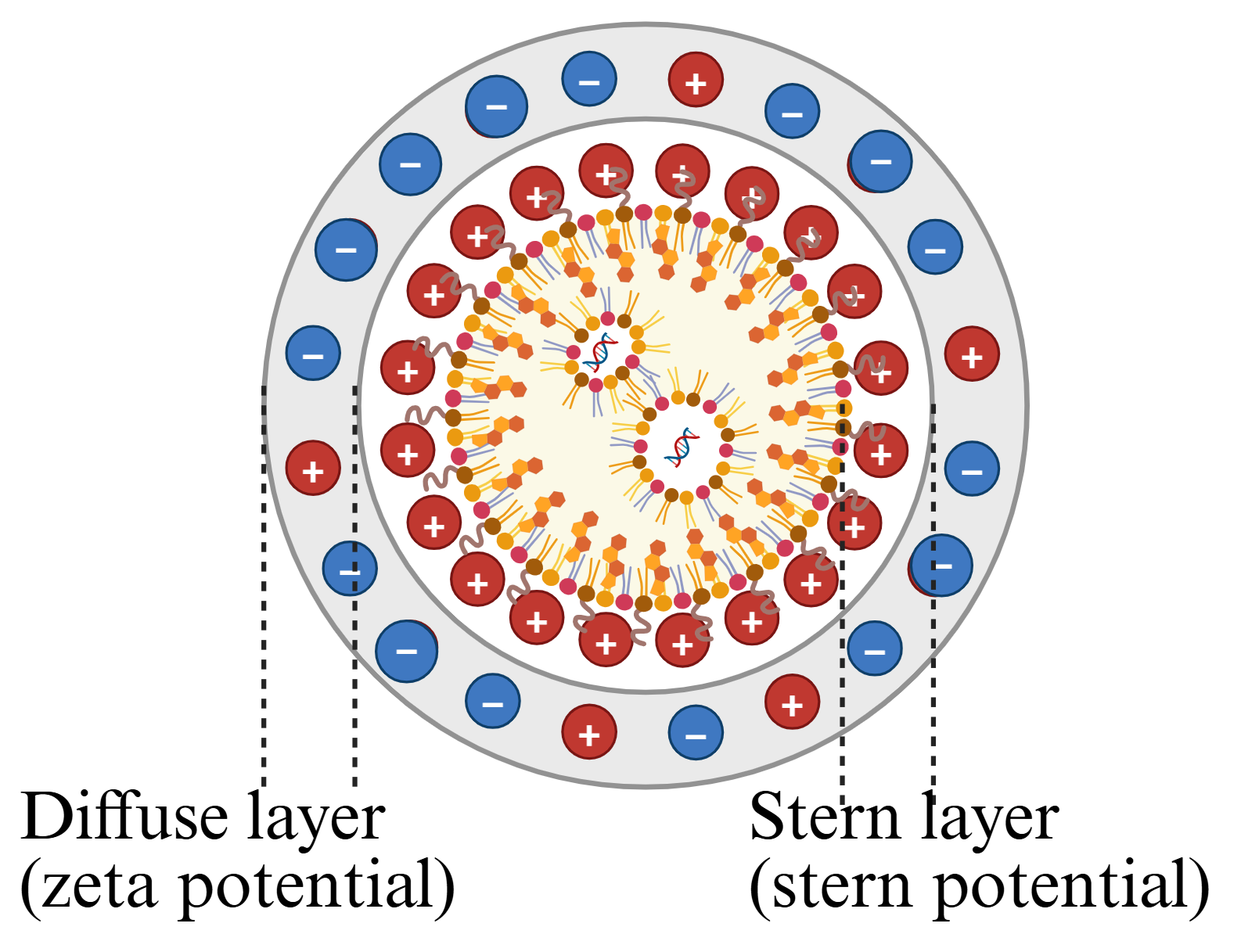}
\caption{The potential of layers of an extracellular vesicle. (Generated by biorender.com)}
\label{fig:Zeta}
\end{figure}

EVs typically carry a net negative surface charge due to the presence of membrane-bound phospholipids, proteins, and glycans \cite{ref3}. When EVs are suspended in a fluid medium, this surface charge interacts with ions in the surrounding environment, forming what is known as the electrical double layer (EDL). The EDL consists of a tightly bound layer of counter-ions at the vesicle surface and a diffuse layer of more loosely associated ions, as shown in Fig. \ref{fig:Zeta}.
Between these layers lies the slipping plane, which defines the boundary at which the fluid begins to flow relative to the moving particle.  

The zeta potential ($\zeta$) is the electrostatic potential at this slipping plane and measures the particle’s surface charge in the context of electrophoresis. It is a key parameter that governs the electrokinetic behaviour of colloidal particles, including EVs. A higher magnitude of zeta potential generally indicates greater electrophoretic mobility.
%\begin{figure}[!t]
%\centering
%\includegraphics[width=3.5in]{Figures/Zeta_potential.png}
%\caption{The potential of layers of a particle \cite{ref1}.}
%\label{fig:Zeta}
%\end{figure}
When an external electric field ($E$) is applied to a fluid containing EVs, the electric force acts on the charged vesicles, causing them to drift \cite{ref27}.
The Smoluchowski equation offers a widely used expression for the electrophoretic velocity ($v_e$) of a spherical particle under these conditions\cite{ref37}
\begin{equation}
 v_e = \frac{\varepsilon \zeta}{\eta}E.
 \label{eqn:simple_eqn}
 \end{equation}
Here, $\varepsilon$ is the permittivity of the fluid, $\eta$ is its dynamic viscosity, and $E$ is the magnitude of the electric field. This equation assumes that the EDL is thin relative to the vesicle radius and that the fluid is in a low-Reynolds-number regime, where viscous forces dominate over inertial effects.

%\subsubsection{Magnetophoretic Drift}
%EVs do not naturally possess magnetic properties and cannot be influenced by magnetic fields without prior modification. It is necessary to implement magnetic characteristics to these vesicles to enable magnetic control over EV motion. This is commonly achieved by incorporating superparamagnetic materials, such as iron oxide nanoparticles (IONPs), into the EV membrane vesicle formation. These magnetically manipulated vesicles retain their biocompatibility and ability to carry molecular cargo while gaining the capacity to respond to external magnetic stimuli\cite{ref15}\cite{ref16}. 
Cho et al. investigated an electrophoretic molecular communication framework that employs a time-varying electric field \cite{cho2022electrophoretic}. Their work explored how the applied field can enhance the number of received particles and reduce the bit error rate in a molecular communication link. In this study, we extend these ideas by focusing on specific electrophoretic features for EVs beam shaping and sweeping over a targeted area.

\subsection{Diffusion and the Fokker-Planck Equation}
The Fokker-Planck equation (FPE) is employed to mathematically describe the evolution of EVs distribution under both random and systematic motion. The FPE models the time evolution of the probability density function (PDF) $P(\mathbf{r}, t)$, representing the likelihood of finding a particle at position $\mathbf{r}$ at time $t$ given by

\begin{equation}
\label{TH-FPE}
\frac{\partial P(\mathbf{r},t)}{\partial t} = -\nabla [\boldsymbol{v_e}(t) P(\mathbf{r},t)] + D\nabla^2P(\mathbf{r},t) + S(\mathbf{r},t),
\end{equation}
 where $D$ is the diffusion coefficient, capturing the strength of random fluctuations given by
\begin{equation}
\label{TH Drift}
D = \frac{k_BT}{6\pi \eta a},
\end{equation}
where $k_B$ is the Boltzmann constant, $T$ is the absolute temperature, $\eta$ is the viscosity of the medium, $a$ is the particle's radius, and $\boldsymbol{r}=({r}_x,{r}_y,{r}_z)$.
 In \eqref{TH-FPE}, $S(\boldsymbol{r},t)$ represents the spatially and temporally distributed source term modelling the EV release process, and $\boldsymbol{v_e}=(v_{e,x},v_{e,y},v_{e,z})$ is a periodic drift velocity describing deterministic motion due to external time harmonic EMF, modelled as
\begin{equation}
\label{TH Drift}
v_{e,i}(t) = v_{0,i}\cos(\omega t+\varphi_i)=\frac{\varepsilon \zeta}{\eta}E_{0,i}\cos(\omega t+\varphi_i), \quad i \in \{x, y, z\},
\end{equation}
where $v_{0,i}$ represents the maximum  drift velocity in the $i$-th direction, capturing the peak strength of the external modulation given by $E_{0,i}$. The variable $\omega$ denotes the angular frequency of the oscillating EMF, defining how rapidly the drift velocity oscillates in time. The phase $\varphi_i$ represents the phase shift specific to each spatial direction, accounting for any initial offset in the temporal oscillation of the drift velocity components. This framework provides a robust theoretical foundation for analyzing how EVs move under a combination of environmental conditions and applied fields.
This form captures the oscillatory nature of externally applied fields, which periodically modulate the direction and magnitude of the drift force acting on the EVs.

\subsection{Analytical Solution to the Fokker–Planck Equation}
To gain insight into how drift and diffusion jointly influence the spatial probability distribution of particles such as EVs, it is useful to seek an analytical solution to the FPE. When the drift term is time-harmonic, the equation becomes analytically intractable in its original form. However, applying an appropriate coordinate transformation can reduce the equation to a solvable form, allowing classical solution techniques such as Green’s functions to be used.
The method starts by transforming the coordinate system into a reference frame with oscillatory drift
\begin{equation}
    {\xi}_i = {r}_i - \int_0^t {v_{e,i}}(\tau) \, d\tau.
\end{equation}
This co-moving frame eliminates the explicit time dependence from the drift term, simplifying the problem. Integrating the drift velocity with respect to time yields the following
\begin{equation}
    \int_0^t v_{e,i}(\tau) \, d\tau = \frac{{v}_{0,i}}{\omega} \sin(\omega t + \varphi) - \frac{{v}_{0,i}}{\omega} \sin(\varphi),
\end{equation}
where $\mathbf{v}_0$ is the amplitude vector of the oscillatory drift, $\omega$ is the angular frequency, and $\boldsymbol{\phi}$ represents the phase vector. Assuming the system is configured such that the coordinate origin is set to eliminate any initial offset, the transformation reduces to
\begin{equation}
   {\xi_i}(t, {r_i}) = {r_i} - \frac{v_{0,i}}{\omega} \sin(\omega t + \varphi_i).
\end{equation}
In this moving frame, the deterministic component of particle motion is absorbed into the coordinate system, leaving a governing equation that reflects only diffusive dynamics.
In the transformed coordinate system, the homogeneous FPE simplifies to
\begin{equation}
    \frac{\partial G(\boldsymbol{\xi}, t)}{\partial t} = D\nabla^2_{\boldsymbol{\xi}}G(\boldsymbol{\xi},t),
\end{equation}
where $\boldsymbol{\xi}=({\xi}_x,{\xi}_y,{\xi}_z)$.
%which is subject to the initial condition
%\begin{equation}
  %  G(\boldsymbol{\xi}, t) = \delta(\boldsymbol{\xi}).
%\end{equation}
The Green’s function solution provides the fundamental solution to the diffusion equation for a point-source initial condition at $S(\boldsymbol{\xi},t) = \delta(\boldsymbol{\xi})\delta(t)$. In three dimensions, this solution takes the form
\begin{equation}
    G(\boldsymbol{\xi}, t) = \frac{1}{(4 \pi D t)^{3/2}} \exp\left(-\frac{\|\boldsymbol{\xi}\|^2}{4Dt}\right).
\end{equation}
To revert to the original laboratory frame, one applies the inverse transformation
\begin{equation}
   {r_i} = {\xi_i} + \frac{{v}_{0,i}}{\omega} \sin(\omega t + {\varphi_i}).
\end{equation}
Substituting back yields the final expression for the probability density function in the original coordinates
\begin{equation}
    G(\boldsymbol{r}, t) = \frac{1}{(4 \pi D t)^{3/2}} \exp\left(-\frac{\|{r_i} - \frac{{v}_{0,i}}{\omega} \sin(\omega t + {\varphi_i})\|^2}{4Dt}\right),
\end{equation}
This solution describes a Gaussian distribution whose center oscillates in time due to the time-dependent drift, while the spread of the distribution increases with time due to diffusion. It provides a theoretical framework for understanding how periodic drift influences the probabilistic transport of particles in a diffusive medium.
%\section{Mathematic setup}
\subsubsection{EV release}
The initial release of EVs into the ECS is modelled as a spatially localized burst, represented by a three-dimensional Gaussian source function. This formulation captures the nature of EV emission from a discrete source region. It assumes a symmetric concentration decay with increasing distance from the release point. The Gaussian profile reflects the finite spatial extent of EV release, consistent with previous models that use such functions to realistically represent the biological spread of EVs from a localized cellular source in the ECM \cite{ref35}\cite{ref36}. The source function is defined as
\begin{equation}
S(\mathbf{r}, t) = \exp\left(-\left\|\frac{r_i - r_{i,L}}{\sqrt2\sigma_i}\right\|^2\right), 
\label{Eq: S3D} 
\end{equation}
where  ${r}_{i,L}$ and $\sigma_i$ represent the center of release and the spatial extent of the burst in $i^{th}$ dimension, repectively. For isotropic emission, the standard deviations are taken to be equal: $\sigma_x = \sigma_y = \sigma_z = \sigma_L$.
\subsubsection{Spatiotemporal Convolution}
The system response to this source must be evaluated to model the full evolution of the EV concentration in space and time. The Green’s function 
$G(\mathbf{r},t)$ for the system has been derived, and the source $S(\mathbf{r},t)$ has been defined; the total solution is obtained through a spatiotemporal convolution of the source function with the Green’s function
\begin{equation}
    P(\mathbf{r},t) = G(\mathbf{r},t) \mathop{\ast} S(\mathbf{r},t),
\label{Eq: Conv}
\end{equation}
where $\mathop{\ast}$ denotes convolution over both space and time. This operation integrates the effect of the initial EV distribution at each point in space and time, accounting for both diffusion and field-driven drift effects governed by the transport model.

\section{Implementation and Results}
This Section presents the results obtained from the simulation framework developed to investigate EV beamforming under ELF EMFs. 
%\begin{figure}[!t]
%\centering
%\includegraphics[width=3.5in]{Figures/Figure 4.png}
%\caption{The simulation scenario consists of a burst release of EVs from a cell. The EVs move in the ECS due to diffusion and drift induced by EFs\cite{ref45}.}
%\label{fig:SetUp}
%\end{figure}
%\subsection{Conceptual and Physical Assumptions}
To create a controllable and focused simulation model, several physical and conceptual assumptions are made regarding the geometry of the ECS, the properties of EVs, and the nature of the applied EMFs.
The ECS is represented as a homogeneous, isotropic, three-dimensional medium without structural heterogeneity. While the true ECS is a highly complex and tortuous environment, this simplification enables a clean analysis of the influence of EMFs on EV transport without introducing additional geometric variability or spatial confinement effects. 

The EVs are modelled as spherical, nanoscale particles released from a localized point source. The release is treated as a burst occurring at time $t = 0$, consistent with sudden secretion events observed in biological systems. It is assumed that the EV population consists of particles that are uniform in size and have the same physical and chemical properties. Specifically, all vesicles are assumed to carry a uniform net negative surface charge due to the assumption that the EDL is thin relative to the vesicle radius and that the surrounding fluid operates under low-Reynolds-number conditions, where viscous forces dominate over inertial effects, enabling the use of the zeta potential to describe electrophoretic mobility. 
%In addition, EVs are assumed to be successfully magnetically functionalized prior to release, allowing them to respond to external magnetic field gradients through magnetophoresis.

The external EMF is simplified to focus on its effect on EV transport. The electric field components $(E_x, E_y, E_z)$ are assumed to be spatially uniform and time-harmonic, with differences only in amplitude and phase between directions. A set of fixed parameters was selected to reflect known biophysical properties and physiologically relevant conditions informed by existing experimental and theoretical literature on EV properties, physiological media, and EMF interactions listed in Table \ref{param1} . 
\begin{table*}[!t]
\caption{Physical parameters and their values used in the simulation to model the geometry of the EVs and their surrounding environment.}
\centering
\begin{tabular}{l|c|c|c|l}
\hline
Parameter             & Symbol     & Value             & Unit            & Reference \\
\hline
EV Radius             & $a$        & $50 \times 10^{-9}$    & $m$             &  \cite{ref39}\\
Viscosity of Water    & $\eta$     & $1 \times 10^{-3}$     & $Pa\cdot s$     &   \\
Permittivity of Water & $\epsilon$ & $7.08 \times 10^{-10}$ & $F/m$           &   \\
Diffusion Coefficient & $D$        & $1 \times 10^{-12}$    & $m^2/s$         &   \\
Zeta Potential        & $\zeta$    & $-20 \times 10^{-3}$   & V               & \cite{ref3} \\
\hline
\end{tabular}
\label{param1}
\end{table*}
\subsection{Analysis of EVs Distribution}
\begin{figure}[!t]
\centering
\subfloat[]{\includegraphics[width=2.5in]{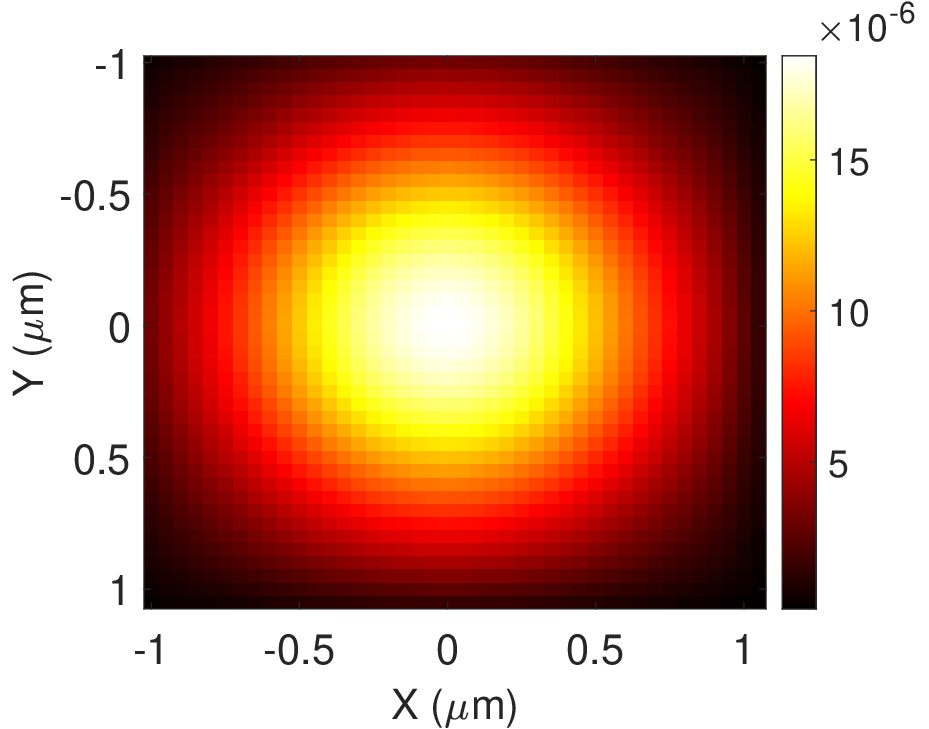}}%
\label{no_emf_1}
\caption{EV concentration distributions considering background noise ($E_{0,x}=E_{0,y}=100V/m$).}
\label{no_emf}
\end{figure}

\begin{figure}[ht]
    \centering
    \begin{minipage}[b]{0.23\textwidth}
        \centering
        \includegraphics[width=\textwidth]{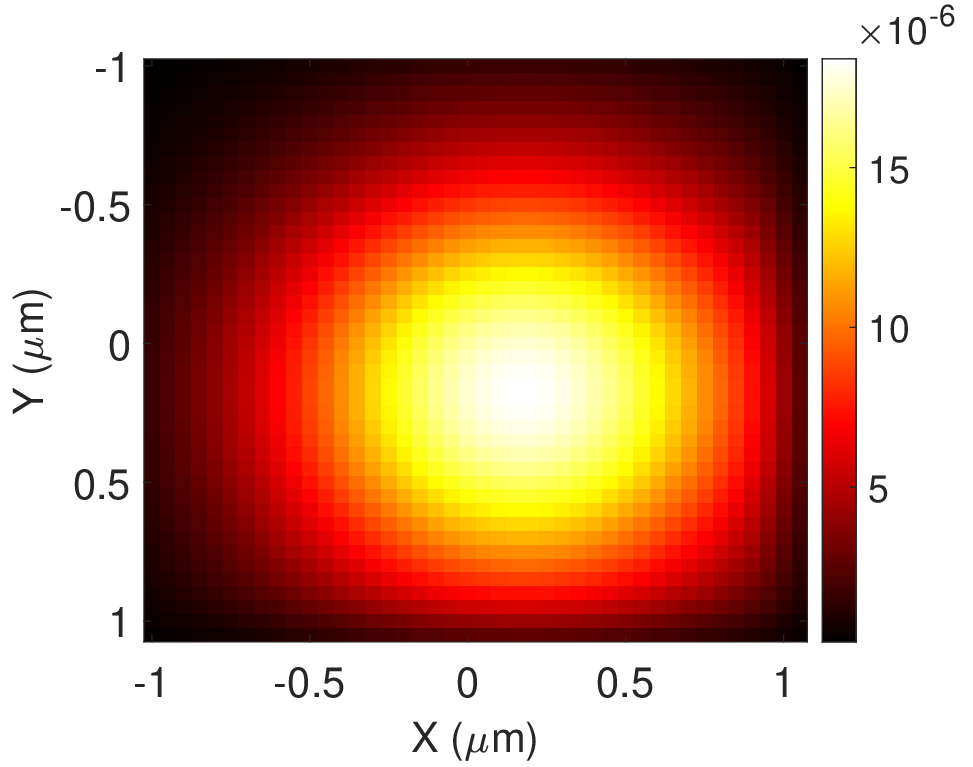}
        \caption*{(a)}
    \end{minipage}\hspace{0.02\textwidth}%
    \begin{minipage}[b]{0.23\textwidth}
        \centering
        \includegraphics[width=\textwidth]{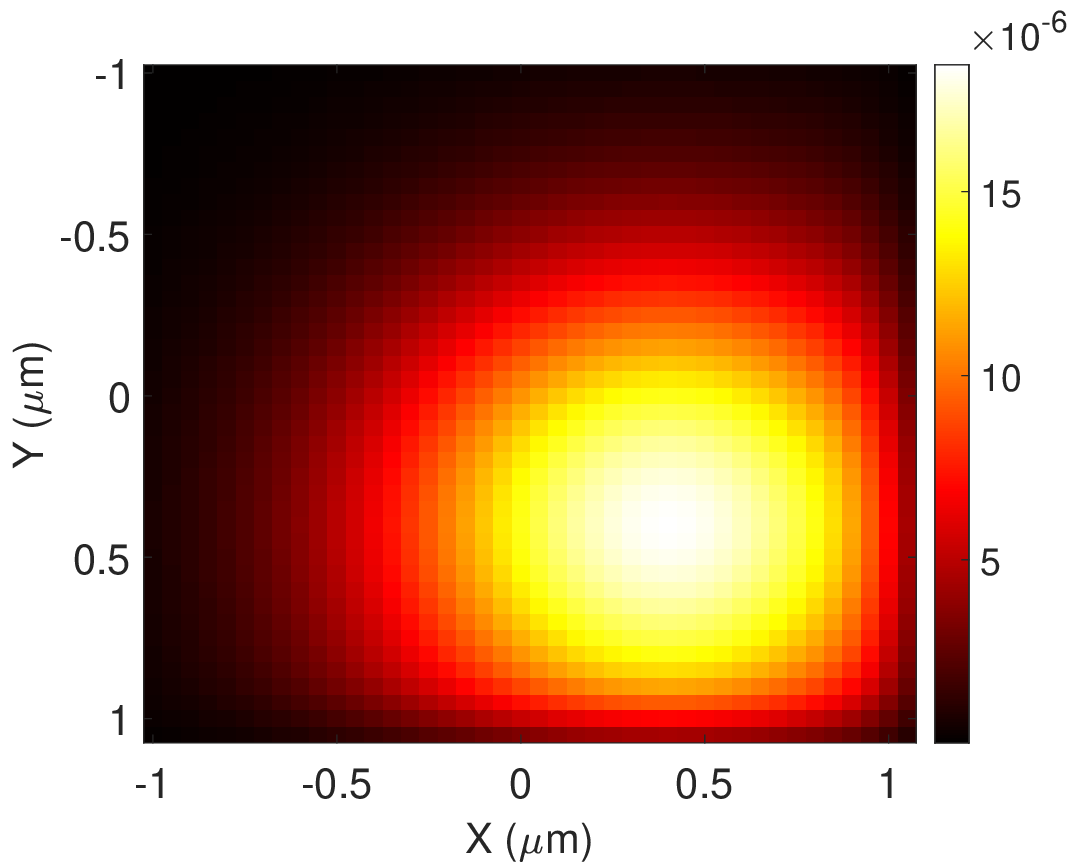}
        \caption*{(b)}
    \end{minipage}\hspace{0.02\textwidth}%
    \begin{minipage}[b]{0.23\textwidth}
        \centering
        \includegraphics[width=\textwidth]{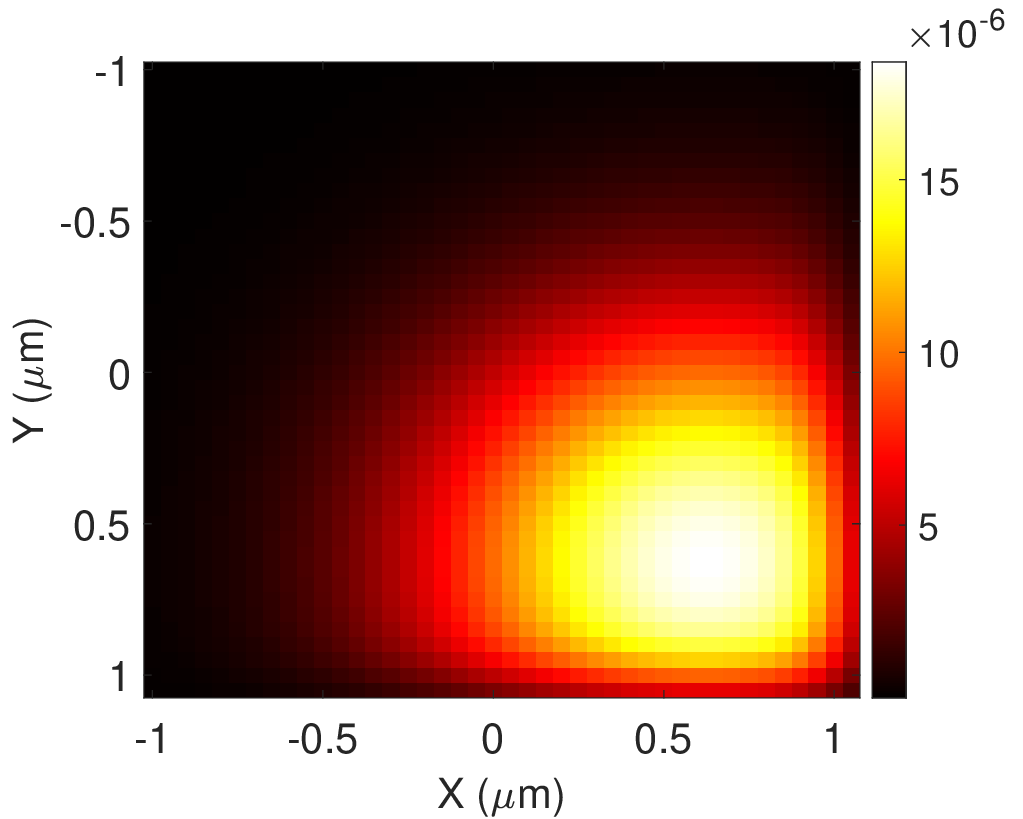}
        \caption*{(c))}
    \end{minipage}\hspace{0.02\textwidth}%
    \begin{minipage}[b]{0.23\textwidth}
        \centering
        \includegraphics[width=\textwidth]{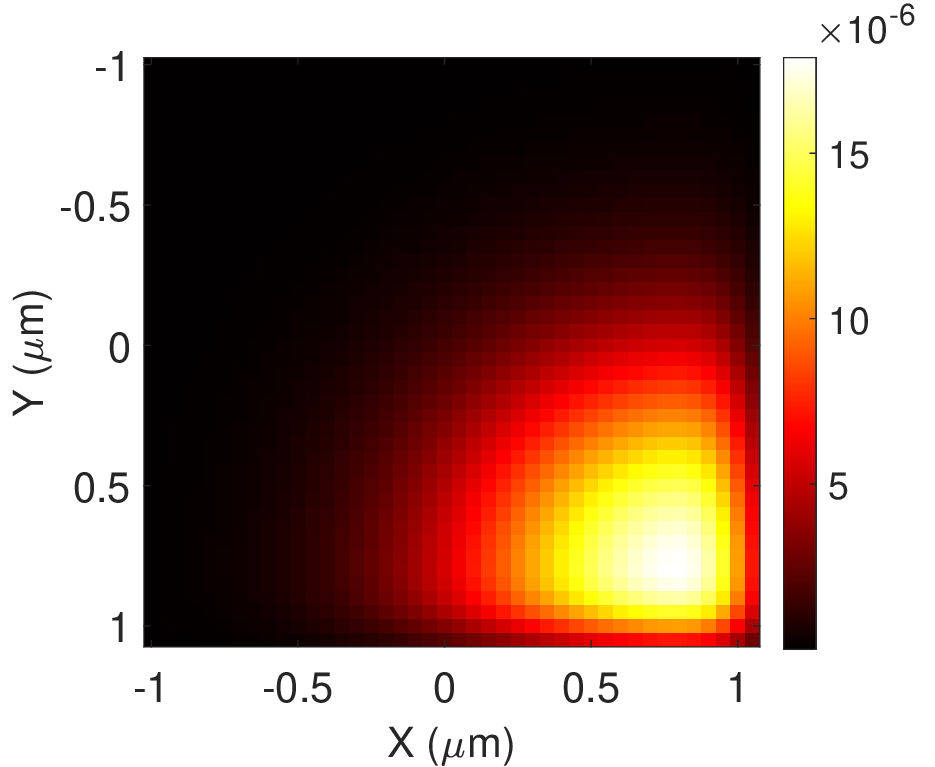}
        \caption*{(d)}
    \end{minipage}
    \caption{Spatial variation of extracellular vesicles beam exposed to electric field of (a) $E=500 V/m$, (b) $E=1000 V/m$, (c) $E=1500 V/m$, and (d) $E=2000 V/m$.}
    \label{fig4}
\end{figure}

 In the absence of EMF ELF (We only consider the background electric field noise of $100 V/m$), the spatial distribution, shown in Fig.~\ref{no_emf}, presents a two-dimensional slice of $P(\mathbf{r}, t)$ through the $x-y$ plane at the midpoint of the $z$-domain, taken at a fixed observation time ($t=150\mathrm{ms}$). As a proof of concept for molecular beamforming, we consider an investigation domain of $2\mu m\times 2\mu m$.  The distribution appears as a symmetric, circular Gaussian shape centred at the origin. 
 %The heatmap illustrates the diffusion of EVs away from the point of release, forming a symmetric profile which spreads out evenly, positively, and negatively over time.
We apply an electric field with four different intensities, 
\( E_{0,x} = E_{0,y} = E=500, 1000, 1500, 2000\ \text{V/m} \), at a frequency of 
\( 5\ \text{Hz} \), with both components having the same phase. 
The corresponding results are presented in Fig.~\ref{fig4}. 
As shown, the EVs are guided in the direction of the applied field, and the 
peak of EV concentration shifts progressively toward the corner of the 
investigation domain as the field intensity increases. However, at higher 
field strengths, we observe a reduction in the peak concentration value 
and the beamwidth of the EV distribution, which can be attributed to EVs 
drifting out of the investigation domain.

\begin{figure}[ht]
    \centering
    \begin{minipage}[b]{0.23\textwidth}
        \centering
        \includegraphics[width=\textwidth]{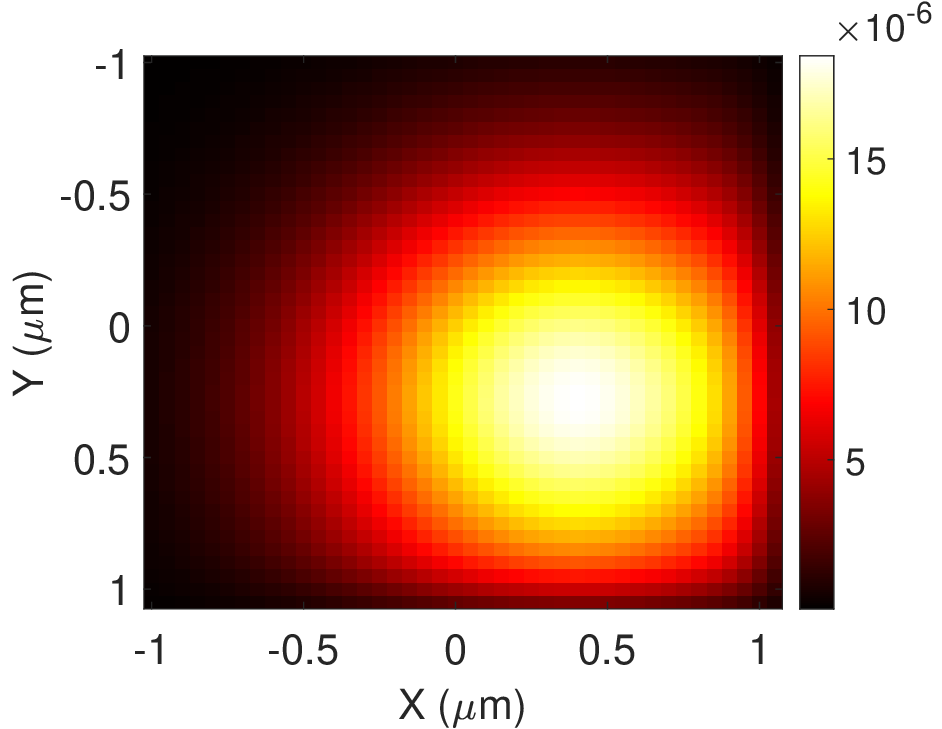}
        \caption*{(a)}
    \end{minipage}\hspace{0.02\textwidth}%
    \begin{minipage}[b]{0.23\textwidth}
        \centering
        \includegraphics[width=\textwidth]{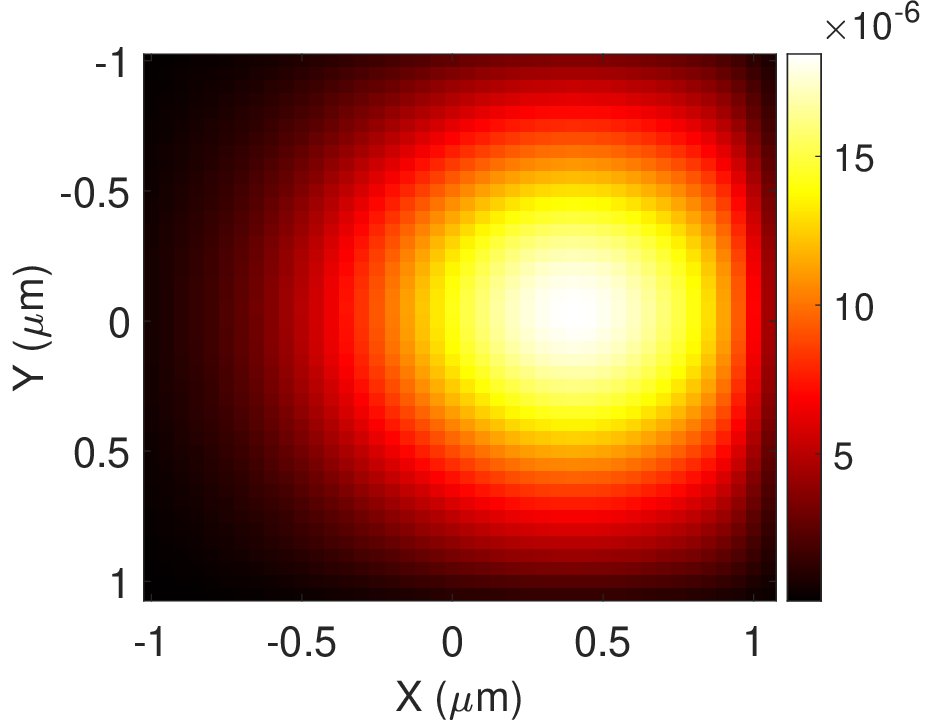}
        \caption*{(b)}
    \end{minipage}\hspace{0.02\textwidth}%
    \begin{minipage}[b]{0.23\textwidth}
        \centering
        \includegraphics[width=\textwidth]{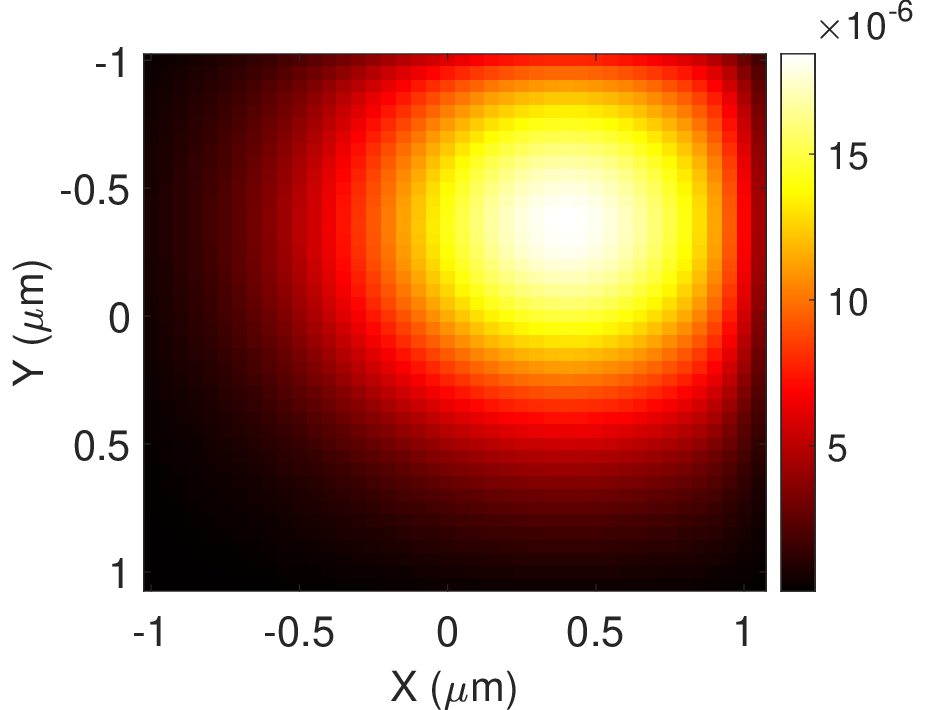}
        \caption*{(c))}
    \end{minipage}\hspace{0.02\textwidth}%
    \begin{minipage}[b]{0.23\textwidth}
        \centering
        \includegraphics[width=\textwidth]{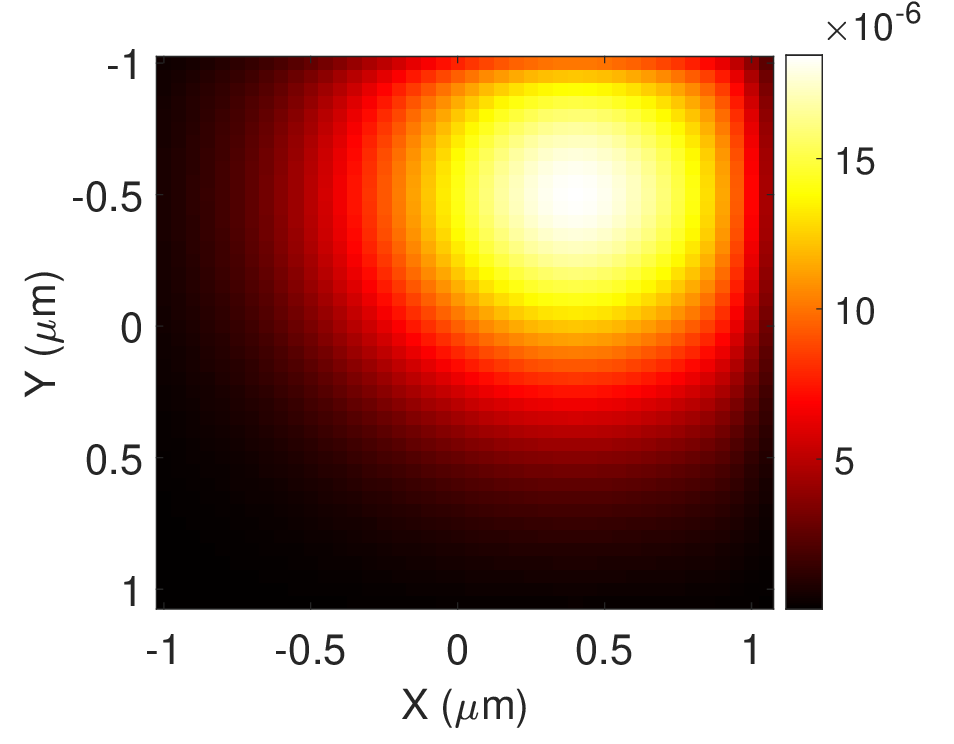}
        \caption*{(d)}
    \end{minipage}
    \caption{Spatial variation of extracellular vesicles beam exposed to electric field with (a) $\varphi_x=45^{\circ}$, (b) $\varphi_x=90^{\circ}$, (c) $\varphi_x=135 ^{\circ}$, and (d) $\varphi_x=180^{\circ}$ having $\varphi_y=0^{\circ}$.}
    \label{fig5}
\end{figure}

To investigate the capability of the electric field in rotating the EV beam, 
we generated Fig. \ref{fig5} depicting the spatial distribution of EVs for 
\( E_{0,x} = E_{0,y} = 1000\ \text{V/m} \) with 
\( \varphi_x = 45^{\circ}, 90^{\circ}, 135^{\circ}, 180^{\circ} \) 
and \( \varphi_y = 0^{\circ} \). 
As shown, a counterclockwise rotation of the beam is observed, while both 
the peak intensity and the beamwidth are preserved.

\begin{figure}[ht]

    \centering
    \begin{minipage}[b]{0.23\textwidth}
        \centering
        \includegraphics[width=\textwidth]{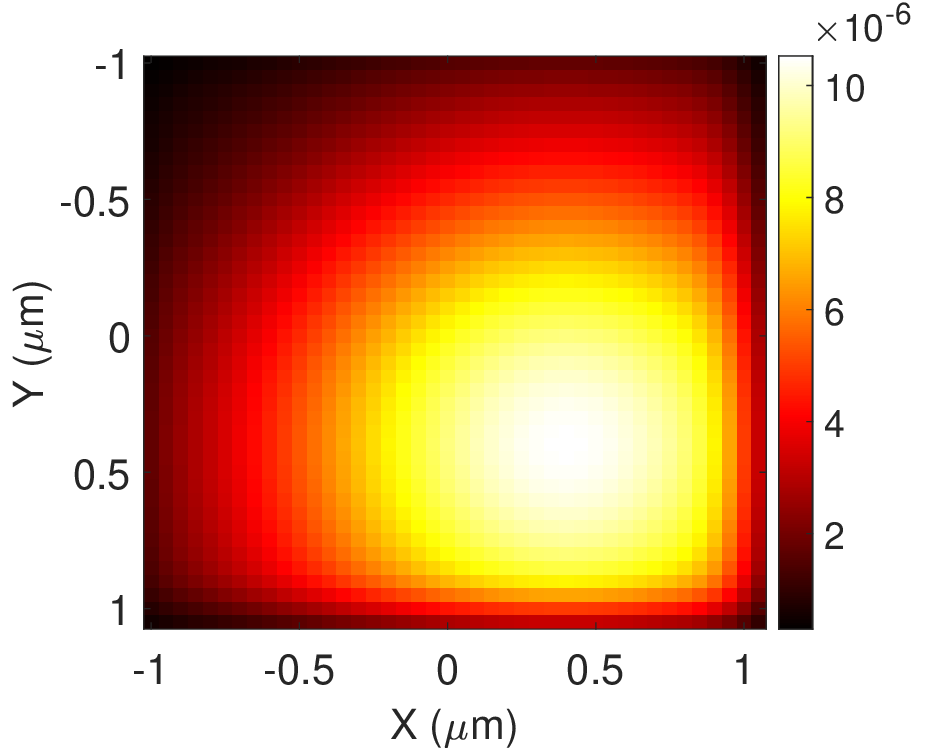}
        \caption*{(a)}
    \end{minipage}\hspace{0.02\textwidth}%
    \begin{minipage}[b]{0.23\textwidth}
        \centering
        \includegraphics[width=\textwidth]{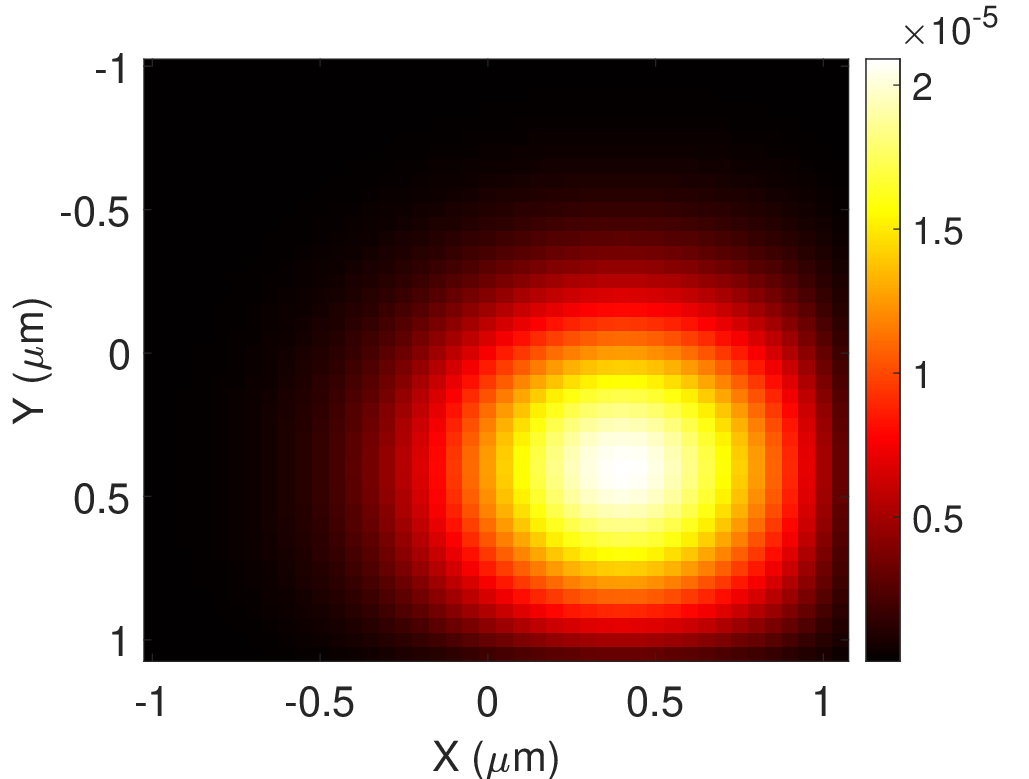}
        \caption*{(b)}
    \end{minipage}\hspace{0.02\textwidth}%
    \begin{minipage}[b]{0.23\textwidth}
        \centering
        \includegraphics[width=\textwidth]{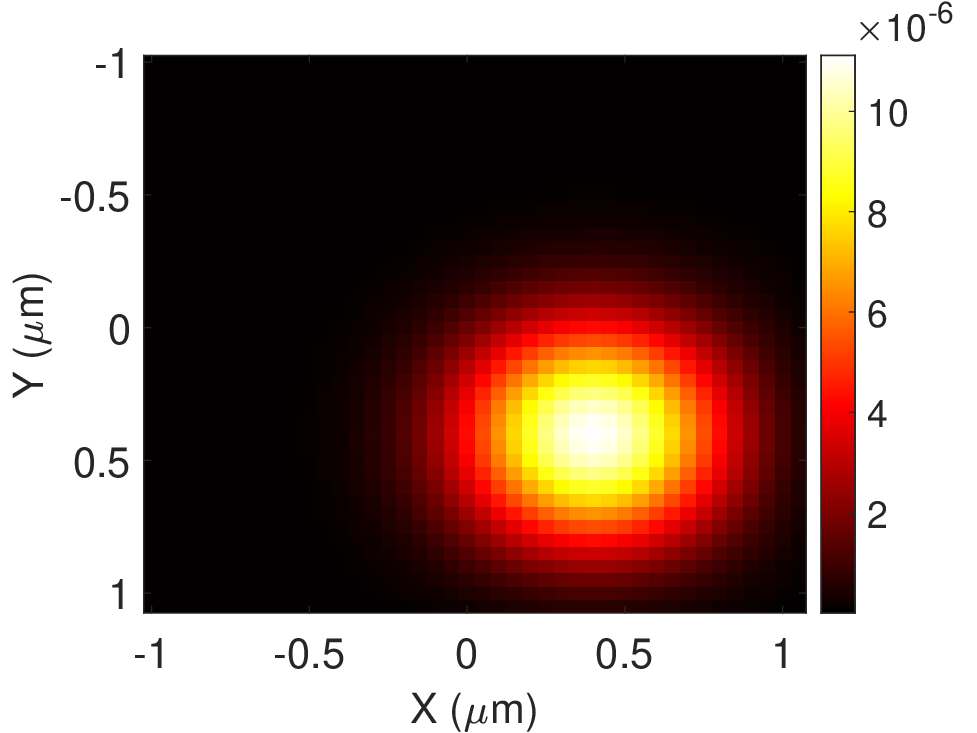}
        \caption*{(c))}
    \end{minipage}\hspace{0.02\textwidth}%
    \begin{minipage}[b]{0.23\textwidth}
        \centering
        \includegraphics[width=\textwidth]{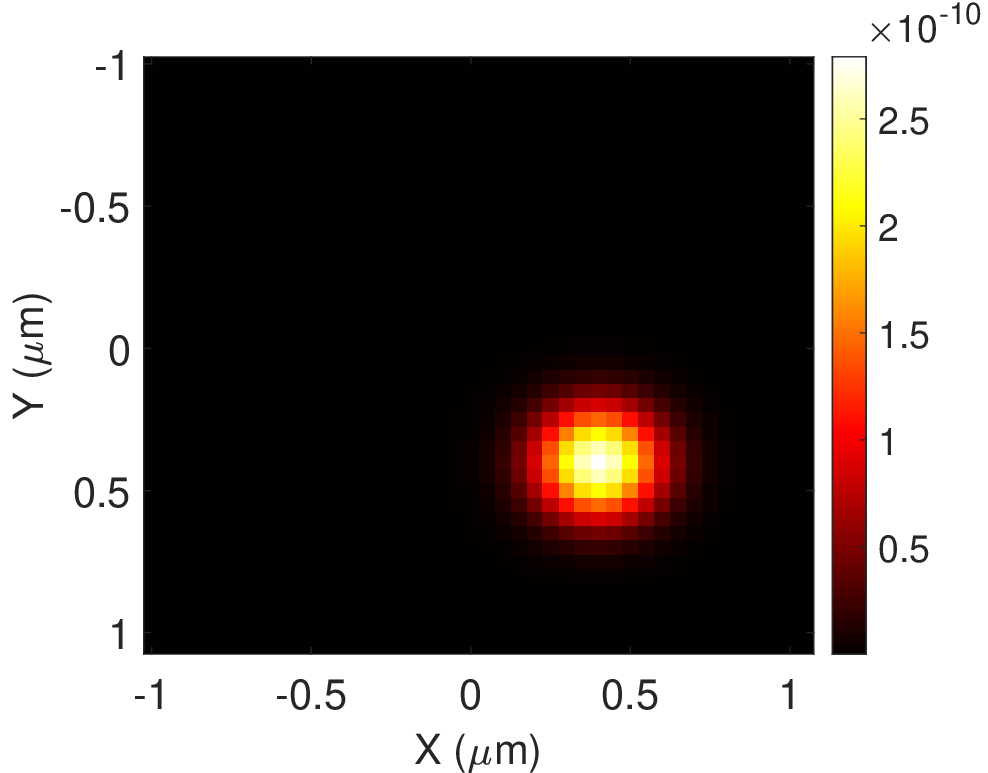}
        \caption*{(d)}
    \end{minipage}
    \caption{Spatial variation of extracellular vesicles beam of size (a) $a=25 nm$, (b) $a=100 nm$, (c) $a=200 nm$, $a=2000 nm$, exposed to electric field of $E_{0,x}=E_{0,y}=1000 V/m$ and zero phases.}
    \label{size}
\end{figure}

It is well established that EVs are secreted with sizes ranging from 
\( 50\ \text{nm} \) (exosomes) to \( 5000\ \text{nm} \) (apoptotic bodies). Fig.~\ref{size} illustrates the spatial distribution of EVs for four different sizes: $a = 25\,\text{nm},\ 100\,\text{nm},\ 200\,\text{nm},\ \text{and}\ 2000\,\text{nm}$.
As shown, the size of EVs has a direct impact on the beamwidth, as larger sizes 
correspond to lower diffusion coefficients, leading to a narrower beam. 
However, the peak value of EV's PDF decreases significantly, since a smaller 
number of larger EVs are released and able to propagate into the environment. 
This illustrates a general principle of molecular beamwidth variation in beamforming, governed by particle size.

%The effects of increasing EFI while maintaining a constant axial ratio are illustrated in Figures~\ref{fig:low_EFI} and~\ref{fig:high_EFI}. These cases correspond to field strengths of $E_x=50 \, V/m$ and $E_x=300 \, V/m$, respectively, allowing an exploration of how increasing field strength influences EV displacement. The spatial distribution in subfigure~\ref{fig:low_EFI_1} exhibits a peak displaced negatively from the source along the $x$-axis, with an almost uniform distribution. The corresponding spatiotemporal plot subfigure~\ref{fig:low_EFI_2} shows a limited negative shift along the $x$-axis with increased dispersion at later times.
%\begin{figure*}[!t]
%\centering
%\subfloat[]{\includegraphics[width=2.5in]{Figures/low_EFI_1.eps}%
%\label{low_EFI_1}}
%\hfil
%\subfloat[]{\includegraphics[width=2.5in]{Figures/low_EFI_2.eps}%
%\label{low_EFI_2}}
%\caption{EV concentration distributions at a low EFI.}
%\label{low_EFI}
%\end{figure*}
%In subfigure~\ref{fig:high_EFI_1}, the spatial distribution shows a peak with significant displacement positively along the $ y$-axis and a very slight negative shift along the $x$-axis, with the peak seemingly out of frame. In subfigure~\ref{fig:high_EFI_2}, the spatiotemporal distribution reveals a slight negative shift along the $x$-axis.
%\begin{figure*}[!t]
%\centering
%\subfloat[]{\includegraphics[width=2.5in]{Figures/high_EFI_1.eps}%
%\label{high_EFI_1}}
%\hfil
%\subfloat[]{\includegraphics[width=2.5in]{Figures/high_EFI_2.eps}%
%\label{high_EFI_2}}
%\caption{EV concentration distributions at a high EFI.}
%\label{high_EFI}
%\end{figure*}

\color{black}

\subsection{Parametric Sensitivity Analysis}
This subsection presents the sensitivity analysis results, examining how the EVs
distribution peak responds to different electric field parameters. We consider 
the electric field amplitude, phase, and frequency as suitable candidates for 
non-invasive, real-time molecular beamforming. The peak position is 
represented in polar coordinates \( (r,\phi) \), describing the radial and 
angular displacement. 

\begin{figure}[!t]
\centering
\includegraphics[width=3.3in]{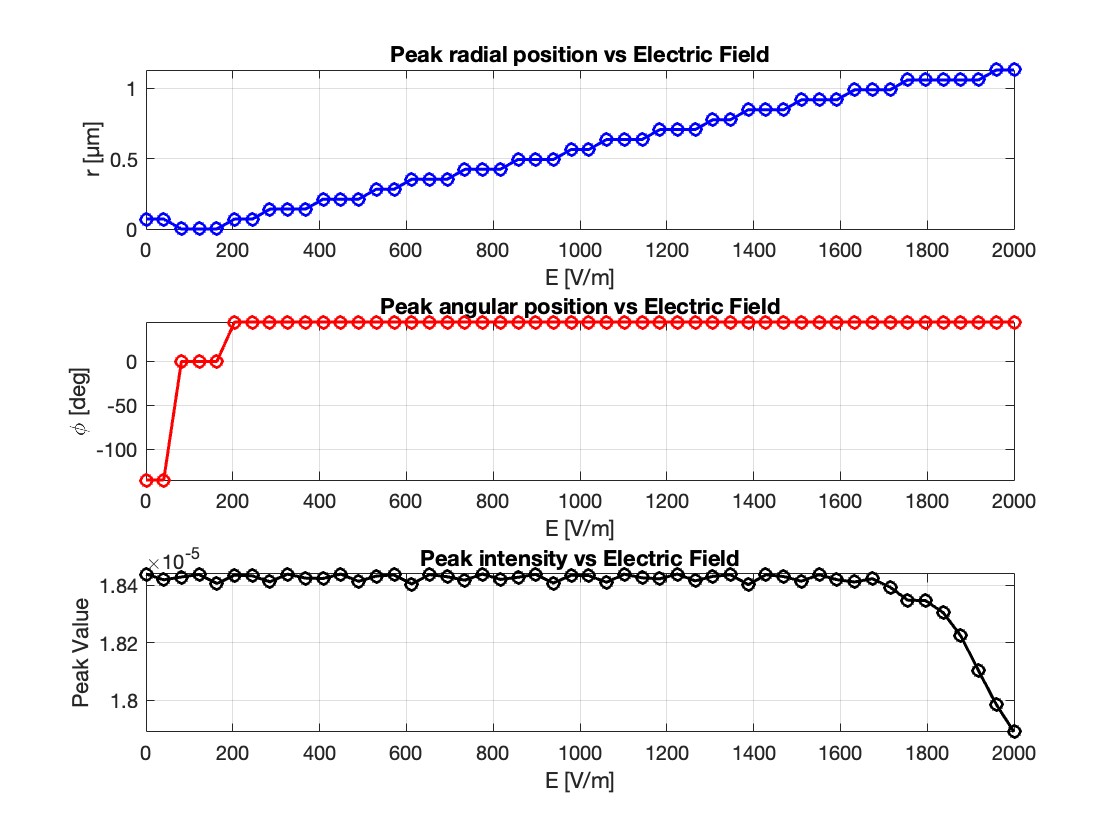}%
\caption{ Radial and angular displacement and value of EVs' PDF peak in terms of electric field intensity.}
\label{fig:EFI}
\end{figure}

Fig.~\ref{fig:EFI} illustrates the displacement of the $P(r_x,r_y,r_z=0, t=150ms)$ peak position
and value as a function of the electric field intensity 
\( E_{0,x} = E_{0,y} = E \, \text{V/m} \). The results clearly indicate that 
\( E = 200\ \text{V/m} \) is the lower threshold for observing a measurable 
drift effect of the field on EVs. Furthermore, the figure shows that the radial 
shift of the EV beam increases with field intensity up to 
\( E \approx 1800\ \text{V/m} \), where the curve reaches a plateau. Beyond this 
point, the peak value begins to decrease, defining the dynamic range of 
electric field–mediated beamforming within the investigated spatial domain. 
It should be noted that \( E = 2000\ \text{V/m} \) remains far below the 
\( 5\ \text{kV/m} \) safety threshold generally considered for avoiding 
lethal effects of electric fields~\cite{ref43}.

\begin{figure}[!t]
\centering
\includegraphics[width=3.3in]{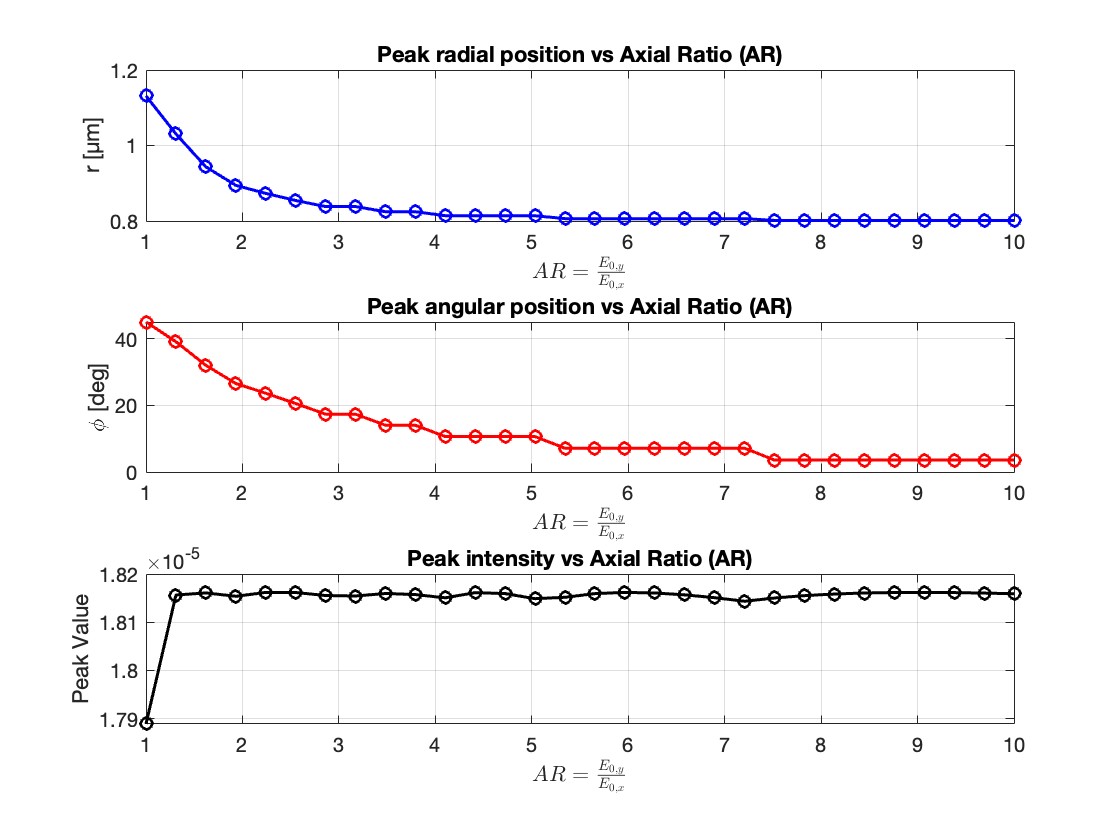}%
\caption{Radial and angular displacement and value of EVs' PDF peak in terms of field axial ratio $(E_{0,y}/E_{0,x})$.}
\label{fig:AR}
\end{figure}

In Fig. \ref{fig:AR} , we explore the impact of the field axial ratio, defined as 
$AR = \tfrac{E_{0,y}}{E_{0,x}}$, on EV beamforming. The figure shows a radial drift of approximately 
$0.3\,\mu\text{m}$ and an angular drift of $45^{\circ}$ for the indicated range of $AR$ variation, 
considering $E_{0,y} = AR \cdot E_{0,x} = 2000$. 
It should be noted that for $AR > 4$ and $AR > 7.5$, the radial and angular drifts respectively 
disappear, which must be taken into account when designing field-mediated EV beam control. 
The beam peak value, however, shows no significant change with respect to $AR$.

\begin{figure}[!t]
\centering
\includegraphics[width=3.3in]{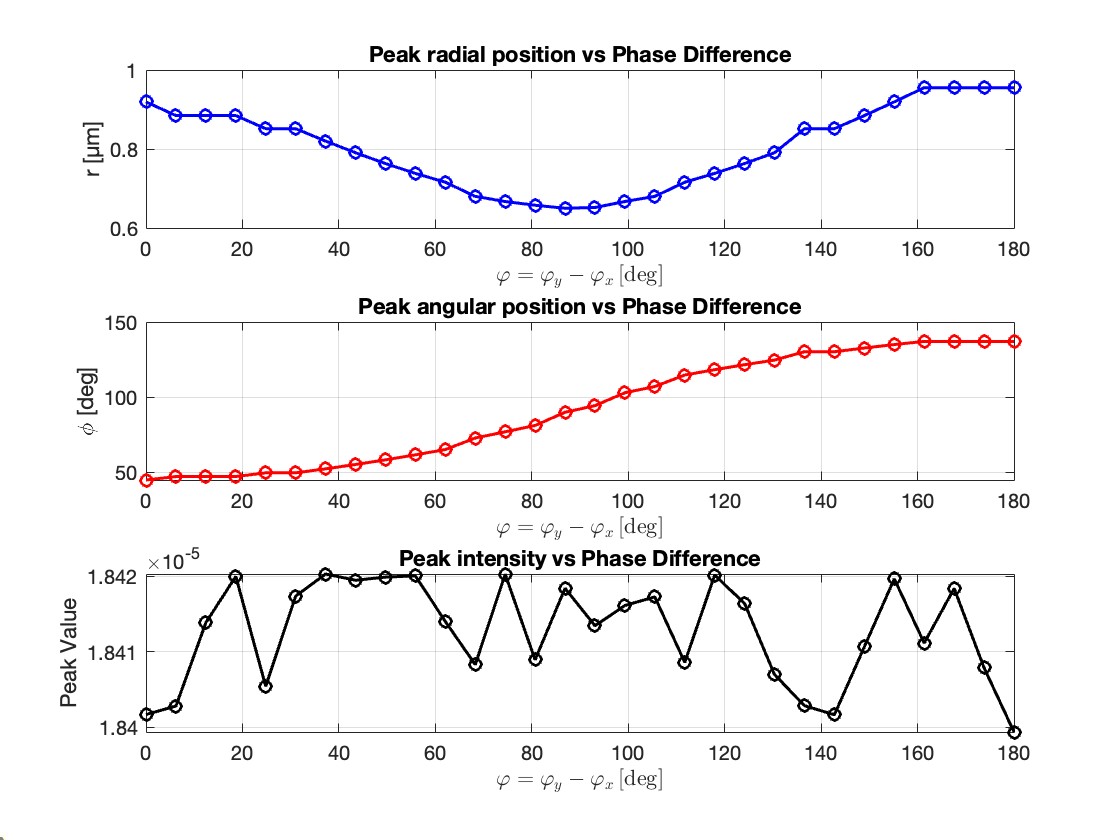}%
\caption{Radial and angular displacement and value of EVs' PDF peak in terms of the phase difference of two field components.}
\label{fig:PD}
\end{figure}

Fig. \ref{fig:PD} illustrates the sensitivity analysis results for varying phase differences $\varphi_y-\varphi_x=\varphi$. The figure shows a beam rotation of up to $100^{\circ}$ for a $180^{\circ}$ phase change, while the peak value remains constant. In this case, the angular drift is more sensitive to field phase variations, particularly around a phase difference of $90^{\circ}$.

\begin{figure}[!t]
\centering
\includegraphics[width=3.3in]{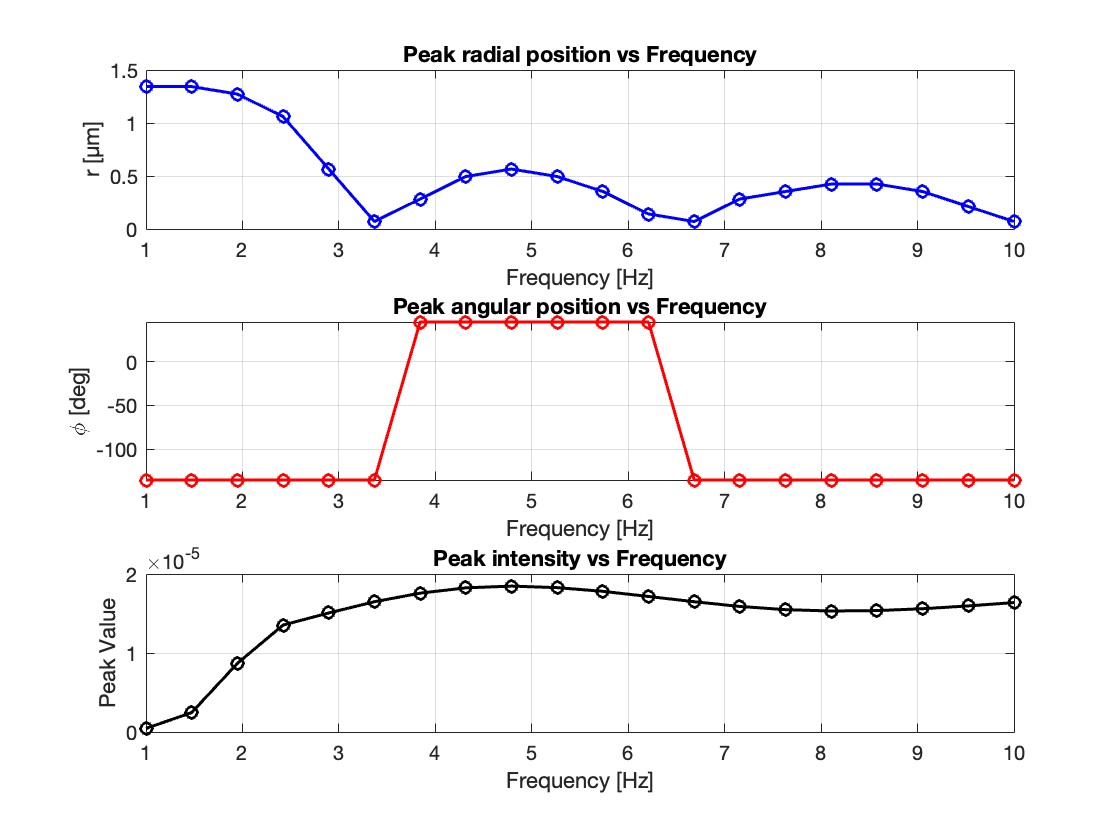}%
\caption{Radial and angular displacement and value of EVs' PDF peak in terms of field's frequency.}
\label{fig:Freq}
\end{figure}

Fig.  \ref{fig:Freq} investigates the effect of frequency on EV displacement. The frequency varied over $f \in [1,10]Hz$, which is within the ELF band relevant for biological field exposure studies\cite{ref9}. The radial drift exhibits a sinc-shaped dependence on frequency, showing only very small beam displacements at high frequencies. This happens because of the dynamics between the oscillating electric field and particle motion. At low frequencies, the field alters direction slowly enough that EVs can respond to the force direction over a longer portion of each cycle. This allows the EVs to build momentum and maintain a net drift in a particular direction. In addition, a null radial drift with angular drift reversal is expected when considering the fixed time point in these studies. Also, a significant peak value is observed for the frequencies larger than $2 Hz$.

\begin{figure}[!t]
\centering
\includegraphics[width=3.3in]{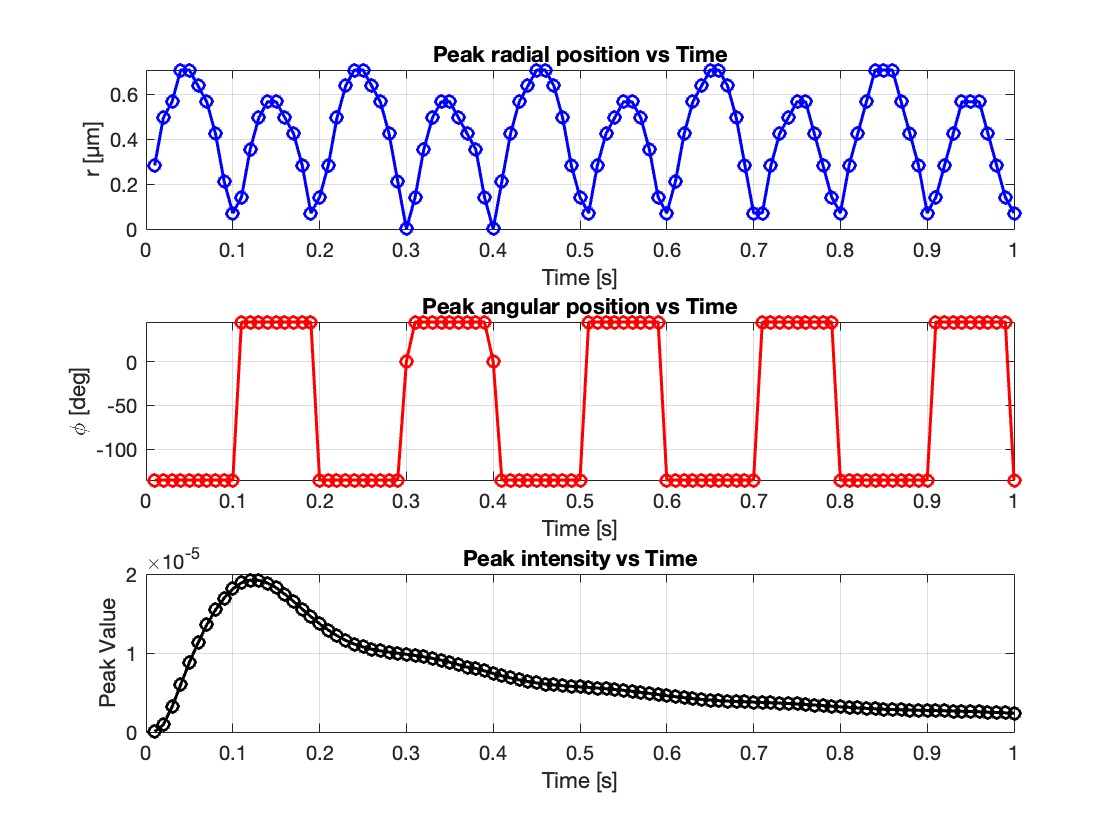}%
\caption{Radial and angular displacement and value of EVs' PDF peak in terms of time for $\varphi=0$.}
\label{fig:Time}
\end{figure}

\begin{figure}[!t]
\centering
\includegraphics[width=3.3in]{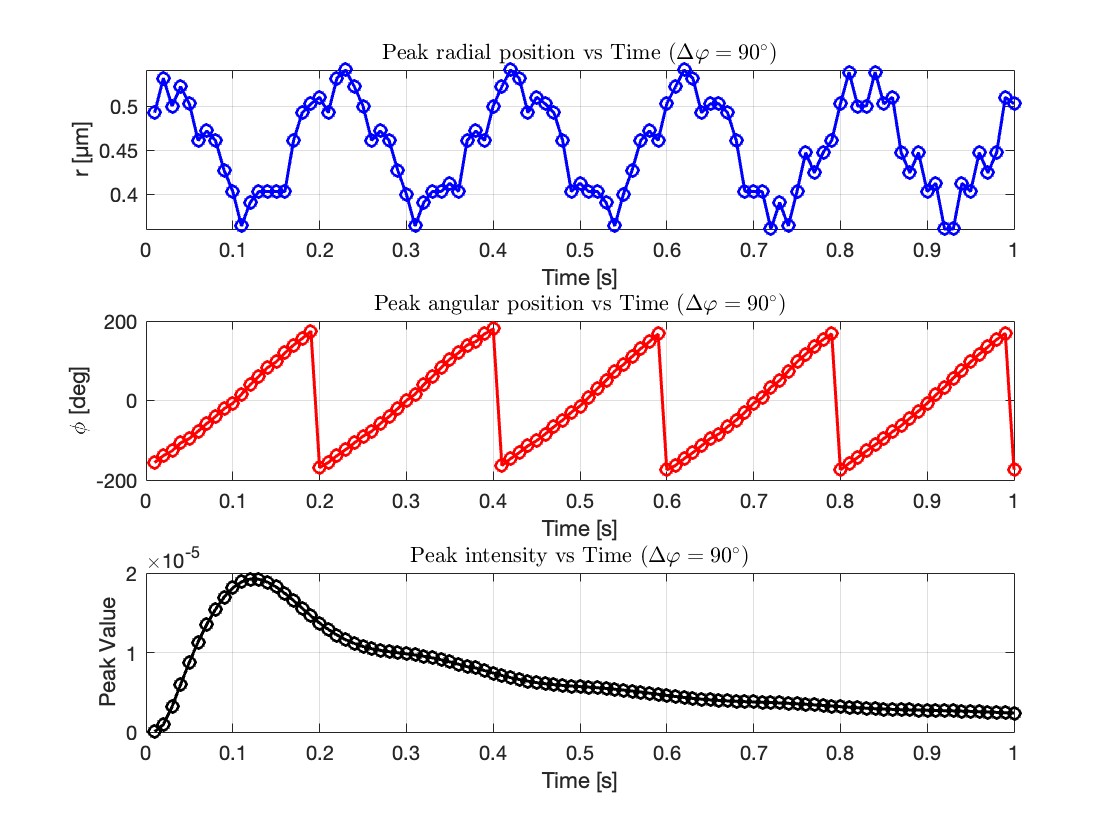}%
\caption{Radial and angular displacement and value of EVs' PDF peak in terms of time for $\varphi=90$.}
\label{fig:Time90}
\end{figure}

Finally, Figs ~\ref{fig:Time} and ~\ref{fig:Time90} present the results of the time-dependent sensitivity analysis, 
highlighting the dynamic behaviour of EV drift under an oscillating electric field with 
$E_{0,x} = E_{0,y} = E \,\text{V/m}$. These Figures display the results for 
$\varphi = 0^{\circ}$ and $\varphi= 90^{\circ}$, respectively. 
As shown, the zero phase difference leads to a periodic radial movement over time with a frequency 
of $5\,\text{Hz}$, as expected. However, the peak value reaches its maximum at $t = 50\,\text{ms}$ 
and decays afterwards. This behaviour can be attributed to the diffusion of particles over time, 
which clearly demonstrates beam degradation. In contrast, a $90^{\circ}$ phase difference between the 
two components of the electric field causes beam rotation with a small radial movement over time, while showing the same peak 
value behaviour.

\begin{table*}[]
\centering
\caption[Sensitivity Analysis Simulation Parameters]{Electric field parameters and ranges used in  sensitivity analysis of EVs drift.}
\label{tab:Swep1}
\begin{tabular}{l|c|c|c|c}
\hline
Parameter        & Symbol            & Value             & Range          & Unit            \\
\hline
Electric Field Intensity  & $E$             & $1000$ & $0-2000$ & $V/m$ \\
Axial Ratio   & $AR$ & $1$  & $1-10$ & $1$             \\
Phase Difference & $\varphi$     & $0$    & $0 - 180$ & $^o$ \\
Frequency        & $\omega$          & $2\pi\times 5$     & $0-10$ & $rad/s$     \\
Time             & $t$               & $150$    & $0-1000$ & $ms$ \\
\hline
\end{tabular}
\end{table*}

\subsection{Biological implications of the results}
The findings of this study provide a promising foundation for using externally applied ELF EMFs to modulate the transport of EVs within a typical ECS. The results demonstrate that controlled directional drift of EVs can be achieved by carefully tuning the electric field parameters. This has profound implications for targeted therapeutic delivery, particularly in the context of chemotherapy.

A key insight from this study is that within the investigated domain of EV beamforming, it is possible to position the beam at an arbitrary location while preserving the peak particle concentration, simply by adjusting the field parameters. The study also identifies clear operating windows in which electric field parameters—such as intensity, frequency, and phase—can achieve effective drift while minimizing energy expenditure and potential tissue heating, which is essential for practical applications in human tissue.
For example, the study revealed that within an electric field intensity (EFI) range of up to $2000,\mathrm{V/m}$, a noticeable directional movement in EV distributions occurs. This indicates that moderate field strengths are sufficient to induce meaningful EV transport, potentially reducing the need for excessively high fields that could raise safety or regulatory concerns. Furthermore, radial drift of the beam can be realized through frequency modulation rather than field intensity, providing an energy-efficient alternative. Low frequencies are particularly valuable, as they align with the safest operating ranges for ELF EMFs.
Another critical observation is the choice of field triggering time point. Without external steering, EVs rely solely on diffusion—a slow and non-directional process in the crowded ECS. Particle diffusion governs the temporal evolution of the molecular beam peak value curve, indicating the specific time point at which maximum beam concentration is achieved.

\section{Conclusion}
This paper investigates the concept of electrophoretic-mediated molecular beamforming 
and its potential application to drug delivery in brain tumors. As a proof of concept, we explore the use of extremely low-frequency electromagnetic fields (ELF EMFs) to non-invasively guide extracellular vesicles (EVs). A theoretical model based on the Fokker--Planck equation is employed to study the effect of a time-harmonic electric field of up to $2000 \,\text{V/m}$ with frequencies below $5 \,\text{Hz}$. Beam positioning is achieved by tuning the field’s intensity, phase, and frequency. 
This work represents an initial demonstration of externally controlled molecular beamforming, with many open directions for future research. The concept could be extended to modulate 
collections of cells or nanomachines toward specific targets with spatial multiplexing in the framework of molecular communication. 
For enhanced performance, engineered EVs with higher zeta potential could be used to strengthen field interactions and increase drift range. Moreover, alternative multiphysics 
approaches such as magnetophoretic beamforming with magnetic nanoparticles may be employed as substitutes for electrophoretic control. Finally, the method also shows potential for 
\textit{in vitro} experiments, enabling the steering of nanoparticles across heterogeneous cell lines toward multiple targets in a switchable manner, while requiring lower field intensities compared to \textit{in vivo} conditions.

%\begin{thebibliography}{44}
% \bibliographystyle{elsarticle-num} 
% \bibliography{name}

\begin{thebibliography}{10}
\expandafter\ifx\csname url\endcsname\relax
  \def\url#1{\texttt{#1}}\fi
\expandafter\ifx\csname urlprefix\endcsname\relax\def\urlprefix{URL }\fi
\expandafter\ifx\csname href\endcsname\relax
  \def\href#1#2{#2} \def\path#1{#1}\fi

\bibitem{gomez2025communicating}
J.~T. G{\'o}mez, P.~Hofmann, L.~Y. Debus, O.~T. Ba{\c{s}}aran, S.~Lotter, R.~Khanzadeh, S.~Angerbauer, B.~D. Unluturk, S.~Abadal, W.~Haselmayr, et~al., Communicating smartly in molecular communication environments: Neural networks in the internet of bio-nano things, arXiv preprint arXiv:2506.20589 (2025).

\bibitem{ref31}
E.~Kianfar, Magnetic nanoparticles in targeted drug delivery: a review, Journal of Superconductivity and Novel Magnetism 34~(7) (2021) 1709--1735.

\bibitem{10149035}
M.~Zoofaghari, F.~Pappalardo, M.~Damrath, I.~Balasingham, Modeling extracellular vesicles-mediated interactions of cells in the tumor microenvironment, IEEE Transactions on NanoBioscience 23~(1) (2024) 71--80.
\newblock \href {https://doi.org/10.1109/TNB.2023.3284090} {\path{doi:10.1109/TNB.2023.3284090}}.

\bibitem{10922788}
M.~Zoofaghari, K.~Sagini, M.~Damrath, A.~Zargarnia, H.~Flaten, M.~Veletić, A.~Llorente, I.~Balasingham, In silico study of bloodstream penetrating extracellular vesicles, IEEE Transactions on Molecular, Biological, and Multi-Scale Communications 11~(2) (2025) 166--175.
\newblock \href {https://doi.org/10.1109/TMBMC.2025.3550323} {\path{doi:10.1109/TMBMC.2025.3550323}}.

\bibitem{9446681}
H.~Arjmandi, H.~K. Rudsari, J.~Santos, M.~Zoofaghari, O.~Ievglevskyi, M.~Kanada, A.~Khaleghi, I.~Balasingham, M.~Veletić, Extracellular vesicle-mediated communication nanonetworks: Opportunities and challenges, IEEE Communications Magazine 59~(5) (2021) 68--73.
\newblock \href {https://doi.org/10.1109/MCOM.001.2000994} {\path{doi:10.1109/MCOM.001.2000994}}.

\bibitem{10032172}
H.~K. Rudsari, M.~Zoofaghari, M.~Damrath, M.~Veletić, J.~Bergsland, I.~Balasingham, Anomalous diffusion of extracellular vesicles in an extracellular matrix for molecular communication, IEEE Transactions on Molecular, Biological, and Multi-Scale Communications 9~(1) (2023) 8--12.
\newblock \href {https://doi.org/10.1109/TMBMC.2023.3240928} {\path{doi:10.1109/TMBMC.2023.3240928}}.

\bibitem{ref11}
E.~H. Apu, et~al., Biomedical applications of multifunctional magnetoelectric nanoparticles, Materials Chemistry Frontiers 6~(11) (2022) 1368--1390.

\bibitem{ref45}
Scientific image and illustration software | biorender, Online, available: https://biorender.com/.

\bibitem{wang2023frequency}
Y.~Wang, G.~A. Worrell, H.-L. Wang, It is the frequency that matters: Effects of electromagnetic fields on the release and content of extracellular vesicles, bioRxiv (2023).

\bibitem{wong2022brief}
C.~J.~K. Wong, Y.~K. Tai, J.~L.~Y. Yap, C.~H.~H. Fong, L.~S.~W. Loo, M.~Kukumberg, J.~Fr{\"o}hlich, S.~Zhang, J.~Z. Li, J.-W. Wang, et~al., Brief exposure to directionally-specific pulsed electromagnetic fields stimulates extracellular vesicle release and is antagonized by streptomycin: A potential regenerative medicine and food industry paradigm, Biomaterials 287 (2022) 121658.

\bibitem{rudsari-2022}
H.~K. Rudsari, M.~Zoofaghari, M.~Veletić, J.~Bergsland, I.~Balasingham, \href{https://doi.org/10.1109/tnb.2022.3206908}{{The End-to-End molecular communication model of Extracellular Vesicle-Based Drug Delivery}}, IEEE Transactions on NanoBioscience 22~(3) (2022) 498--510.
\newblock \href {https://doi.org/10.1109/tnb.2022.3206908} {\path{doi:10.1109/tnb.2022.3206908}}.
\newline\urlprefix\url{https://doi.org/10.1109/tnb.2022.3206908}

\bibitem{wang2024engineering}
S.~Wang, Y.~Mao, S.~Rong, G.~Liu, Y.~Cao, Z.~Yang, H.~Yu, X.~Zhang, H.~Fang, Z.~Cai, et~al., Engineering magnetic extracellular vesicles mimetics for enhanced targeting chemodynamic therapy to overcome ovary cancer, ACS Applied Materials \& Interfaces 16~(30) (2024) 39021--39034.

\bibitem{ref3}
G.~Midekessa, et~al., Zeta potential of extracellular vesicles: Toward understanding the attributes that determine colloidal stability, ACS Omega 5~(27) (2020) 16701--16710.

\bibitem{ref27}
L.~S. Cheung, S.~Sahloul, A.~Orozaliev, Y.-A. Song, Rapid detection and trapping of extracellular vesicles by electrokinetic concentration for liquid biopsy on chip, Micromachines 9~(6) (2018) 306.

\bibitem{ref37}
J.~Lim, et~al., Magnetophoresis of nanoparticles, ACS Nano 5~(1) (2010) 217--226.

\bibitem{cho2022electrophoretic}
S.~Cho, T.~C. Sykes, J.~P. Coon, A.~A. Castrej{\'o}n-Pita, Electrophoretic molecular communication with time-varying electric fields, Nano Communication Networks 31 (2022) 100381.

\bibitem{ref35}
H.~K. Rudsari, et~al., The end-to-end molecular communication model of extracellular vesicle-based drug delivery, IEEE Transactions on NanoBioscience 22~(3) (2022) 498--510.

\bibitem{ref36}
G.~Wilemski, On the derivation of smoluchowski equations with corrections in the classical theory of brownian motion, Journal of Statistical Physics 14~(2) (1976) 153--169.

\bibitem{ref39}
L.~M. Doyle, M.~Z. Wang, Overview of extracellular vesicles, their origin, composition, purpose, and methods for exosome isolation and analysis, Cells 8~(7) (2019) 727.

\bibitem{ref43}
S.~Wang, J.~Sun, R.~M. Dastgheyb, Z.~Li, Tumor-derived extracellular vesicles modulate innate immune responses to affect tumor progression, Frontiers in Immunology 13 (Nov. 2022).

\bibitem{ref9}
Z.-C. Sun, et~al., Extremely low frequency electromagnetic fields facilitate vesicle endocytosis by increasing presynaptic calcium channel expression at a central synapse, Scientific Reports 6~(1) (Feb. 2016).

\end{thebibliography}

\vfill
\end{document}